\documentclass[preprint]{elsarticle}

\usepackage{graphicx}
\usepackage{dcolumn}
\usepackage{bm}
\usepackage{natbib}
\usepackage{color}
\usepackage{hyperref}
\usepackage{mathtools}

\begin{document}


\title{Landauer-B\"uttiker conductivity for spatially-dependent uniaxial strained armchair-terminated graphene nanoribbons}

\author[AEC]{Abdiel E. Champo}
\address[AEC]{
Facultad de Ciencias en F\'isica y Matem\'aticas, Universidad Aut\'onoma de Chiapas, Carretera Emiliano Zapata, Km. 8, Rancho San Francisco, 29050, \\Tuxtla Guti\'errez, Chiapas, M\'exico.}

\author[PRT]{Pedro Roman-Taboada}
\address[PRT]{Instituto de Fisica, Universidad Nacional Aut\'onoma de M\'exico, \\Apartado Postal 20-364,01000,Ciudad de M\'exico, M\'exico.}

\author[PRT]{Gerardo G. Naumis\corref{cor1}}
\ead{naumis@fisica.unam.mx}

\date{\today}

\begin{abstract}
The Landauer-B\"uttiker conductivity of arbitrary uniaxial spatially dependent strain in an armchair graphene nanoribbon is studied. Due to the uniaxial character of the strain, the corresponding transfer matrix can be reduced to a product of $2\times2$ matrices. Then the conductivity and the Fano factor can be calculated from this product. As an example of the technique, sinusoidal space dependent strain fields are studied using two different strain wavelengths. For the bigger wavelength the conductivity is reduced when compared with the unstrained case, although both conductivities are almost the same in shape. Whereas, for the smaller wavelength case, the conductivity is strongly modified. In spite of this, for energies close to the Dirac point energy, the conductivity and the Fano factor are quite similar to their unstrained counterpart for the two strain wavelengths here studied.
\end{abstract}

\maketitle


\section{Introduction}

Graphene, the first truly two dimensional (2D) material experimentally growth \cite{castro2009theelectronic,Review10}, has  very interesting and fascinating transport properties \cite{Roche08,torres2014introduction}. In particular, when a deformation field is applied to it, novel effects arise \cite{carrillo2014gaussian,carrillo2016strained,ChenSi16,naumis2017electronic,Akinwande17}. For instance, one can mention the well-known gap opening  in a graphene nanoribbon when an uniaxial and uniform strain field is applied to it \cite{Pereira09}. This fact opened a new field of research known as straintronics, which aims to fine tune the electronical and optical properties by the application of mechanical deformations \cite{naumis2014mapping}. Following this direction, many theoretical works studying the effect of mechanical deformations in the transport properties of graphene have been made \cite{naumis2009design,diaz2016self-similar,garcia2016angle-dependent,garcia2017self}. For example, the transport properties of graphene nanoribbons connected to metallic leads have been studied using a tight-binding approach \cite{zhang2010crossover,zhang2011dependence,shi2012exact,gao2014electronic}, as well as using {\it ab initio} calculations \cite{barraza2010effects,barraza2012charge,barraza2013coherent}. These kind of studies have been done for both the unstrained and strained cases \cite{jing2015mechanical,zenan2015coupling}. 

For the unstrained case, perhaps the most relevant discovery is the role played by the topology of the ribbon, that is, the form of the ribbon's edges. In fact, it has been shown that, in the limit of large width-to-length ratio and low energy, armchair-terminated graphene nanoribbons (AGNs) have a finite conductivity at the Dirac point, whereas zigzag terminated graphene nanoribbons (ZGNs) have a vanishing conductivity at the same point \cite{shi2012exact}. Another important fact is the presence of an asymmetry in the conductivity of AGNs. This phenomenon is related to an electron-hole asymmetry and to the use of metallic leads \cite{zhang2011dependence}. It is important to say that such asymmetry is not seen in the case of ZGNs  since the edges of a ZGN protect the electron-hole symmetry \cite{zhang2011dependence}. On the other hand, for the case of uniform strained AGNs, it was found that the conductivity can be suppressed with respect to the pristine case if the strain amplitudes are larger than a threshold \cite{jing2015mechanical}, needless to say that the asymmetry seen in the unstrained case still appears in the strained case. In the case of an uniform strained ZGN, it has been found that the conductivity has two sharp peaks around the van Hove singularities. Note that the van Hove singularities are not longer at energy $E=\pm t_0$ ($t_0$ being the hopping parameter for pristine graphene) as in graphene, but they are shifted from their original positions due to the strain field, see \cite{Pereira09,naumis2017electronic} for more details. This fact can be understood by an effective decoupling of the ZGN into weakly coupled dimmers \cite{naumis2014mapping,naumis2017electronic}.

Even though the cases of uniform strained ZGNs and AGNs have attracted a lot of attention and are well understood \cite{topsakal2010current-voltage,oliva2014anisotropic,jing2015mechanical,zenan2015coupling,maurice2017low-energy}, to our best knowledge, the effects of non-uniform strain fields in the transport properties still deserve research. Therefore, in this paper, we study the effects of non-uniform  uniaxial strain fields in the transport properties of armchair graphene nanoribbons. To this end, we follow the formalism developed in reference \cite{shi2012exact}, which is based on the transfer method approach using a tight-binding Hamiltonian. Then we generalize their method to an uniaxial space-dependent strained AGN. As will be seen later on, our method works for any kind of uniaxial strain field applied along the armchair direction of the graphene ribbon. When applied to the particular case of a periodic strain, the results show important changes of the conductivity, however, for low energies (this is, energies near the Dirac cone energy), the conductivity is almost unaffected by the strain field. 

The article is organized as follows. In section \ref{Model} we introduce the tight-binding Hamiltonian used to describe the electronic properties of an uniaxial strained armchair-terminated graphene nanoribbon (AGN) connected to two metallic leads, which are considered to be semi-infinite square lattices. Then, in section \ref{transfermatrix} we generalize the transfer matrix method developed in reference \cite{shi2012exact}  to study the transport properties of an AGN under any kind of uniaxial strain field applied along the armchair direction of the graphene ribbon. Section \ref{Results} is devoted to the application of our method to study the transport properties of an AGN under uniaxial spatially periodic strain. Some conclusions are given in section \ref{conclusions}. Finally, in appendices A, B, and C some calculations 
regarding the transfer matrix method here developed are presented.

\section{Uniaxial strained armchair-terminated graphene ribbon}
\label{Model}

\begin{figure*}
    \centering
    \includegraphics[scale=0.424]{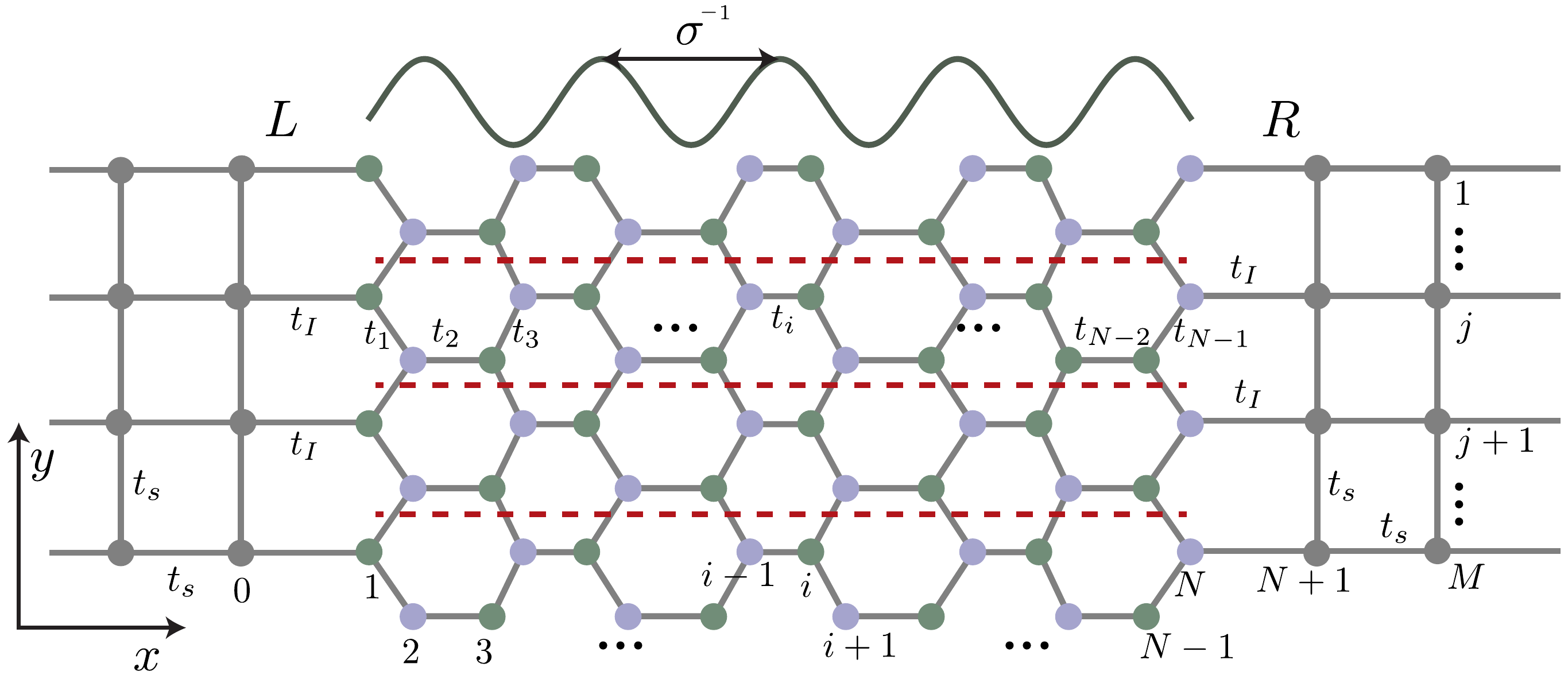}
    \caption{Schematic representation of an armchair graphene nanoribbon connected to electrodes at interfaces L (left) and R (right). The wavy line on top of the nanoribbon indicates a strain applied in the $x$ direction. For a sinusoidal strain field, the modulation has wavelength $\sigma^{-1}$. As a result, the distances of carbon atoms are affected, although owing to the uniaxial nature of the strain, columns of atoms are displaced the same.}
    \label{fig:grafeno}
\end{figure*}

In this section we introduce the model to be used for describing the electronic properties of a strained AGN. We start with an AGN joined to two metallic leads as indicated in Fig. \ref{fig:grafeno}. It is important to mention that the metallic leads will be represented by semi-infinite square lattices joined to the AGN at the interface by a hopping parameter given by $t_I$. Then we apply an uniaxial spatially periodic strain field along the armchair direction of the considered ribbon (here on, we will denote the armchair direction as the $x$-direction, see Fig. \ref{fig:grafeno}). We suppose, as a first approximation, that the leads are unaffected by the strain field. Due to the presence of the strain field the atoms are shifted from their original positions in pristine graphene. The new positions of the carbon atoms are given by  $\textbf{r}'_{i,j}= \textbf{r}_{i,j}+\textbf{u}(x_{i},y_{j})$, where $\textbf{r}_{i,j}=(x_{i},y_{j})$ are the unstrained coordinates of the atoms (see Fig. \ref{fig:grafeno}) and $\textbf{u}(x_{i},y_j)$ is the displacement-deformation vector or, in other words, the strain field. The label $i=1,2,..., N$ ($j=1,2,..., M$) indicates the site's position of the carbon atoms along the $x$-direction ($y$-direction), see Fig. \ref{fig:grafeno}.  For an uniaxial strain field along the $x$-direction, in general, there is a deformation along the $y$-direction, which is determined by the Poisson ratio with value $\nu=0.165$ for graphene \cite{jing2015mechanical}. Here we will simplify the calculus by suposing that the strain in the $y$-direction can be neglected, in such a way that symmetry is preserved along the $y$-axis, i.e., we will assume that $\textbf{u}(x_{i},y_{j}) \approx (u(x_{i}),0)$. Notice however that such field can be obtained by choosing a suitable external strain field.

On the other hand, as we proved in a previous work, the electronic properties of an uniaxial strained AGN are well described, in the low energy limit, by the following nearest-neighbor tight-binding Hamiltonian \cite{naumis2014mapping},
\begin{equation}
\hat{H}= - \sum_{\textbf{r}_{i,j}} t_{\textbf{r}'_{i,j} + \delta'_{i,n}} c_{\textbf{r}'_{i,j}}^{\dag}c_{\textbf{r}'_{i,j}+{{\bf{\delta}}'_{i,n}}}+ \epsilon_{0} \sum_{\textbf{r}'_{i,j}} c_{\textbf{r}'_{i,j}}^{\dag}c_{\textbf{r}'_{i,j}}
\label{hatH}
\end{equation}
where $\textbf{r}'_{i,j}$ runs over all sites of the strained lattice of the AGN, $\epsilon_{0}$ is the onsite energy corresponding to sites of the deformed lattice. In other words, $\epsilon_0$ is the chemical potential and can be adjusted by an effective gate voltage applied directly to the graphene ribbon \cite{gao2014electronic}. The operator $c_{\textbf{r}'_{i,j}}^{\dag}$ $\left(c_{\textbf{r}'_{i,j}+ {\bf{\delta}}'_{i,n}}\right)$ creates (annihilates) an electron in a carbon atom at the position $\textbf{r}'_{i,j}$ $\left(\textbf{r}'_{i,j}+ {\bf{\delta}}'_{i,n}\right)$ in the strained AGN. Finally, ${\bf{\delta}}'_{i,n}$ (with $n=1,2,3$) are the vectors that point to the three nearest neighbors of a carbon atom at the site $\textbf{r}'_{i,j}$ in the strained lattice. For unstrained graphene, ${\bf{\delta}}'_{i,n}= {\bf{\delta}}_{n}$ with,
\begin{equation}
\begin{array}{lcc}
\delta_{1}= (-1)^{i+1}\left(1, \sqrt{3}\right)a/2 \\
 \delta_{2}= (-1)^{i+1}\left(1, -\sqrt{3}\right)a/2\\
  \delta_{3}=(-1)^{i+1}\left(-1,0\right)a,
\end{array}
\end{equation}
where $a=0.145\text{ nm}$ is the interatomic distance in pristine graphene. The hopping parameters in a strained AGN depend upon the position of the carbon atoms since the strain field induces changes in the orbitals' overlap. For an uniaxial strain field, the hopping parameters, in the small strain's amplitude limit (this is for strain's amplitudes much lower than the lattice constant of pristine graphene), are given by \cite{naumis2014mapping},
\begin{equation}
    t_{\textbf{r}'_{i,j}+ \delta'_{i,n}}=t_G \exp{\left[-\beta \left(-1+| \delta'_{i,n} |/a \right)\right]},
    \label{hoppinggeneral}
\end{equation}
where $\beta \approx 3.37$ is the Gr\"uneisen parameter. $t_G \approx 2{.}7$eV is the hopping integral for pristine graphene. For the sake of simplicity, we will measure all distances in unit of $a$, this is equivalent to take $a=1$. The quantity $| \delta'_{i,n} |$ is the distance between carbon atoms in pristine AGN and its strained counterpart. For the small strain's amplitude limit, we have \cite{naumis2014mapping},
\begin{equation}
 \delta'_{i,n}=|\delta_n|\cdot \left[\mathbf{u}(\mathbf{r}+\mathbf{\delta}_n)-\mathbf{u}(\mathbf{r})\right].
\end{equation}
%

\section{Transfer matrices and transmission coefficient}
\label{transfermatrix}

This section is devoted to develop the transfer matrix method that will be used to study the transport properties of a strained AGN. We have basically generalized the method obtained in reference \cite{shi2012exact}. Following reference \cite{shi2012exact}, we start by stating the Schr\"odinger equation to be solved,
\begin{equation}
    \hat{H} \Psi(E)= E \Psi(E),
    \label{schro}
\end{equation}
where $\hat{H}$ is the Hamiltonian defined in Eq. (\ref{hatH}) that describes the electronic properties of a strained AGN. $\Psi(E)$ is the wave function of the system at a given energy $E$. In the Landauer-B\"uttiker formalism the wave function of the system can be represented as,
\begin{equation}
    \Psi(E)= \sum_{\textbf{r}'_{i,j}} \alpha_{i,j} \vert \psi_{i,j} \rangle,
    \label{wavefunction}
\end{equation}
where the complex coefficients $\alpha_{i,j}$ can be determined using the matrix transfer method. Then, if $E_{F}$ is the Fermi energy of the electrodes, which is set by their occupation $n_{c}$, there must be $M$ right-traveling and $M$ left-traveling waves (channels) in each electrode. Each of these channels is characterized by a transverse wave-number $k_{y}^{s}= \frac{ \pi s}{M+1}$ with $s=1,\ldots,M$ and by a longitudinal wave-number $k_{x}^{s}$ related to $k_{y}^{s}$ by the dispersion relation of the leads, this is, by the dispersion relation of a semi-infinite square lattice, 
\begin{equation}
    E_{F}=-2t_{S}(\cos k_{x}^{s} + \cos k_{y}^{s}).
\end{equation}

On the other hand, it is useful to write the Schr\"odinger equation in terms of the wave function Eq. (\ref{wavefunction}). To do that we substitute the wave function Eq. (\ref{wavefunction}) into Eq. (\ref{schro}). After some manipulations, one gets,
\begin{equation} \label{alphas}
-\sum_{\tau, \gamma } t_{\textbf{r}'_{i,j}+ \delta'_{i,n}} \alpha_{i+\tau, j+\delta}= (E- \epsilon_{0}) \alpha_{i,j},
\end{equation}
where $\tau$ and $\gamma$ specify the nearest-neighboring sites of $\textbf{r}'_{i,j}$. Observe that for uniaxial strain applied along the armchair direction, the translation symmetry along the $y$-direction (or, in other words, the zigzag direction) is not broken, therefore we have $t_{\textbf{r}'_{i,j}+ \delta'_{i,n}}=t_{\textbf{r}'_{i,l}+ \delta'_{i,n}}$ for all $j,l=1,\ldots,M$ and $|\delta'_{i,1}|=|\delta'_{i,2}|$. Using the previous fact and Eq. (\ref{hoppinggeneral}) one obtains that $t_{\textbf{r}'_{i,j}+ \delta'_{i,1}}=t_{\textbf{r}'_{i,j}+ \delta'_{i,2}}$. It follows that for each column $i$ there are only two different values of the hopping parameters which join different columns, this fact allows us to define $t_i$ as the hopping integral between a site in the column $i$ and its nearest neighbor in the column $i+1$, as shown in Fig. \ref{fig:grafeno}. Until now we have followed the method developed in Ref. \cite{shi2012exact}. From here on, we proceed to generalize their method for the case of an AGN under an uniaxial strain field applied along the $x$-direction. For that end we start by observing that for non-uniform strain there are four different cases for Eq. (\ref{alphas}), namely,
\begin{equation}
(E- \epsilon_{0})\alpha_{i,j}= \left \lbrace \begin{array}{lcc}
-[t_{i} (\alpha_{i+1,j}+ \alpha_{i+1,j-1})+ t_{i-1} \alpha_{i-1,j}]    
\\ -[t_{i}\alpha_{i+1,j}+ t_{i-1}(\alpha_{i-1,j}+\alpha_{i-1,j+1})]     
\\ -[t_{i}(\alpha_{i+1,j}+\alpha_{i+1,j+1})+t_{i-1}\alpha_{i-1,j}]      
\\ -[t_{i}\alpha_{i+1,j}+ t_{i-1}(\alpha_{i-1,j}+\alpha_{i-1,j-1})],  
\end{array} \right.
\label{alphafour}
\end{equation}
where each row of the previous expression is applied to $i \equiv 0,1,2,3(\mod{4})$, starting from top to bottom. For example, for the first four columns of graphene (see Fig. \ref{fig:grafeno}):
\begin{equation}
 \begin{array}{lcc}
(E- \epsilon_{0})\alpha_{1,j}=-[t_{1} (\alpha_{2,j}+ \alpha_{2,j-1})+ \tilde{t}_{0} \alpha_{0,j}]  \\
\\ (E- \epsilon_{0})\alpha_{2,j}=-[t_{2}\alpha_{3,j}+ t_{1}(\alpha_{1,j}+\alpha_{1,j+1})]  \\
\\(E- \epsilon_{0})\alpha_{3,j}= -[t_{3}(\alpha_{4,j}+\alpha_{4,j+1})+t_{2}\alpha_{2,j}]  \\
\\ (E- \epsilon_{0})\alpha_{4,j}=-[t_{4}\alpha_{5,j}+ t_{3}(\alpha_{3,j}+\alpha_{3,j-1})].
\end{array} 
\end{equation}
The next step is to express the $M$ coefficients $\alpha_{i,j}$ for a given $i$ as the vector
\begin{equation}
    \vec{\alpha}_{i}= \left( \begin{array}{lcc}
         \alpha_{i,1} \\
         \vdots \\
   \\ \alpha_{i,M} 
    \end{array} \right).
    \label{alphavec}
\end{equation}
Now we define the $2M\times2M$ transfer matrix  $\hat{P}_i$, which connects $\vec{\alpha_{i}}$ with its neighboring columns, as follows,
\begin{equation} \label{ec4}
\left(\begin{array}{lcc}
     \vec{\alpha}_{i-1}  \\
   \\  \vec{\alpha}_{i}
\end{array}\right)= \hat{P}_{i} \left(\begin{array}{lcc}
     \vec{\alpha}_{i}  \\
  \\   \vec{\alpha}_{i+1}
\end{array}\right).
\end{equation}

Observe that there must be four different kinds of transfer matrices $\hat{P}_i$ because we have four inequivalent atoms per each row of the AGN. It follows that the system can be seen as made of blocks of four columns, where each block changes in the $x$-direction. For a system with $N$ columns (this is, $N$ atoms along the $x$-direction per row), we will have $N/4$ blocks. We will numerate each of these blocks using the label $q(r)$, such that $q(r)= 4(r-1)+1$ with $r=1,2,3,\ldots, N/4$. By numerating the columns in this way, we are able to define four different transfer matrices, thus reducing the problem of finding $N$ transfer matrices to just find four, as was done in reference \cite{shi2012exact}, but for a general uniaxial strained AGN. We define the transfer matrix that connects the column at the site $q(r)$ with its counterpart at the site $q(r)+1$ as $\mathbf{A}_{r}$. In a similar way, we have $\mathbf{B}_{r}$ (which connects the column at $q(r)+1$ with the one at $q(r)+2$), $\mathbf{C}_{r}$ (which connects the column at $q(r)+2$ with the one at $q(r)+3$), and $\mathbf{D}_{r}$ (which connects the column at $q(r)+3$ with the one at $q(r)+4$). 
As detailed in appendix A, such matrices are given by,
\begin{equation} \label{matricesA}
\textbf{A}_{r}= \left( \begin{array}{lcc}
    -\frac{(E-\epsilon_{0})}{t_{q_{(r)}-1}} \mathcal{I}_{M \times M} & -\frac{t_{q_{(r)}}}{t_{q_{(r)}-1}} \textbf{X}  \\
     \\   \mathcal{I}_{M \times M} & \textbf{0} 
\end{array}\right)
\end{equation}
\begin{equation} \label{matricesB}
\textbf{B}_{r}= \left( \begin{array}{lcc}
    -\frac{(E-\epsilon_{0})}{t_{q_{(r)}}} \textbf{Y}^{t} & -\frac{t_{q_{(r)}+1}}{t_{q_{(r)}}} \textbf{Y}^{t}  \\
     \\   \mathcal{I}_{M \times M} & \textbf{0} 
\end{array}\right)
\end{equation}
\begin{equation} \label{matricesC}
\textbf{C}_{r}= \left( \begin{array}{lcc}
    -\frac{(E-\epsilon_{0})}{t_{q_{(r)}+1}} \mathcal{I}_{M \times M} & -\frac{t_{q_{(r)}+2}}{t_{q_{(r)}+1}} \textbf{X}^{t}  \\
     \\   \mathcal{I}_{M \times M} & \textbf{0} 
\end{array}\right)
\end{equation}
\begin{equation} \label{matricesD}
\textbf{D}_{r}= \left( \begin{array}{lcc}
    -\frac{(E-\epsilon_{0})}{t_{q_{(r)}+2}} \textbf{Y} & -\frac{t_{q_{(r)}+3}}{t_{q_{(r)}+2}} \textbf{Y}  \\
     \\   \mathcal{I}_{M \times M} & \textbf{0} 
\end{array}\right)
\end{equation}
where  $\mathcal{I}_{M \times M}$ is the $M\times M$ identity matrix. $\mathbf{X}$ is a $M\times M$ upper bidiagonal matrix with non-zero elements equal to one and $\mathbf{Y}=\mathbf{X}^{-1}$.

Then the total transfer matrix of our system $\hat{P}$ defined as,
\begin{equation} \label{matrizdetrnsferencia1}
\left( \begin{array}{lcc}
     \vec{\alpha}_{0}^{L}  \\
    \\ \vec{\alpha}_{1}^{L}
 \end{array} \right) =  \hat{P} \left( \begin{array}{lcc}
     \vec{\alpha}_{N}^{R}  \\
   \\  \vec{\alpha}_{N+1}^{R}
 \end{array} \right),
\end{equation}
is given by,  
\begin{equation}
\begin{aligned}
\hat{P}=&\left( \begin{array}{lcc}
    {\tilde{t}_{2}} \mathcal{I}_{M \times M} & \textbf{0} \\
     \\ \textbf{0} &  \mathcal{I}_{M \times M} 
\end{array}\right)\times\\
&
\left(\prod_{r=1}^{N/4} \textbf{A}_{r}\textbf{B}_{r}\textbf{C}_{r}\textbf{D}_{r}\right) \times\left( \begin{array}{lcc}
 \mathcal{I}_{M\times M} &  \textbf{0}  \\
     \\  \textbf{0} & \tilde{t}_{1} \mathcal{I}_{M \times M}  
\end{array}\right),
\end{aligned}
\end{equation}
where $\tilde{t}_{2}= t_St_1/t_I^2$ and $\tilde{t}_1=t_I^2/(t_St_{N-1})$.
In addition, we can obtain explicitly the total product of $\textbf{A}_{r}\textbf{B}_{r}\textbf{C}_{r}\textbf{D}_{r}$ for each block $r$ in terms of $\textbf{X}$ and $\textbf{Y}$, as is done in appendix B; this is useful since that allows us to reduce the problem to a system of bilinear equations.

Now suppose that we have M right-traveling waves and M left-traveling waves for a given energy with unitary amplitude, in the right and left leads, respectively. If this is the case, the right-traveling wave in the $s$-th channel will be scattered into the $s'$-th channel leading to the following wave functions at the edges of the interface between the ribbon and the leads,
\begin{equation}
\begin{array}{lcc}
    \alpha_{ij}^{L}= \sum_{s'}(\delta_{s's}e^{i k_{x}^{s'} x_{i}} + r_{s's}e^{-ik_{x}^{s'}x_{i}}) \sin(k_{y}^{s'}y_{j}) \\
     \\  \alpha_{ij}^{R}= \sum_{s'} t_{s's} e^{i k_{x}^{s'} x_{i}}  \sin(k_{y}^{s'}y_{j}),
\end{array}
\end{equation}
where  $r_{s's}$ and $t_{s's}$ are the reflection and transmission coefficients from the $s$-th to the $s'$-th channel respectively, and $\delta_{s's}$ is the Kronecker delta. $L$ ($R$) stands for the left (right) lead. We can greatly simplify the expression of the wave functions at the edges of the AGN if we rewrite them in terms of the matrices $\hat{\Delta}$ (with size $2M\times M$), $\hat{R}$ (with size $2M\times M$) and $\hat{T}$ (with size $2M\times 1$), which are defined as,
\begin{equation}
\hat{\Delta}= \left( \begin{array}{lcc}
 \hat{\zeta}& 0  \\
    0 & \hat{\zeta} 
\end{array} \right) \left( \begin{array}{lcc}
 \hat{\xi}(x_{0})\\
  \hat{\xi}(x_{1})
\end{array} \right),
\end{equation}
\begin{equation}
\hat{R}= \left( \begin{array}{lcc}
 \hat{\zeta}& 0  \\
    0 & \hat{\zeta} 
\end{array} \right) \left( \begin{array}{lcc}
 \hat{\xi}(x_{0})^{*}\\
  \hat{\xi}(x_{1})^{*}
\end{array} \right),
\label{Rmatrix}
\end{equation}
\begin{equation}
\hat{T}= \left( \begin{array}{lcc}
 \hat{\zeta}& 0  \\
    0 & \hat{\zeta} 
\end{array} \right) \left( \begin{array}{lcc}
 \hat{\xi}(x_{N})\\
  \hat{\xi}(x_{N+1})
\end{array} \right),
\label{Tmatrix}
\end{equation}
with
\begin{equation}
\hat{\zeta}= \left( \begin{array}{lcc}
    \sin(k_{y}^{1} y_{1}) & \ldots & \sin(k_{y}^{M} y_{1})  \\
    \vdots & \vdots & \vdots \\
    \sin(k_{y}^{1}y_{M}) & \ldots & \sin(k_{y}^{M}y_{M})  
\end{array} \right),
\end{equation}
and
\begin{equation}
\hat{\xi}(x_{i})= \left( \begin{array}{lccc}
    e^{k_{x}^{1} x_{i}} & 0 & \ldots & 0  \\
    0 & e^{k_{x}^{2}x_{i}} & \ldots & 0 \\
    \vdots& \vdots & \ddots & \vdots \\
    0 & 0 & \ldots& e^{k_{x}^{M}x_{i}}   
\end{array} \right).
\end{equation}

By using Eqs. (\ref{Rmatrix}), (\ref{Tmatrix}), and the vector notation stablished in Eq. (\ref{alphavec}), it follows that,
\begin{equation} \label{operadorreflexion}
     \left(\begin{array}{lcc}
         \vec{\alpha}_{0}^{L} \\
         \\ \vec{\alpha}_{1}^{L} 
     \end{array} \right)= \hat{R}\hat{r_{s}}+ \hat{\Delta}\hat{\delta_{s}},
\end{equation}
and
\begin{equation} \label{operador transmision}
     \left(\begin{array}{lcc}
         \vec{\alpha}_{N}^{R} \\
         \\ \vec{\alpha}_{N+1}^{R} 
     \end{array} \right)= \hat{T}\hat{t_{s}},
\end{equation}
where 
\begin{equation}
     \hat{r_{s}}= \left( \begin{array}{lcc}
     r_{1,s} \\
     \vdots \\
     r_{M-1,s}\\
     r_{M,s}
     \end{array} \right); \hat{\delta_{s}}= \left( \begin{array}{lcc}
     \delta_{1,s} \\
     \vdots \\
     \delta_{s,s}\\
     \vdots \\
     \delta_{M,s}
     \end{array} \right) ;
     \hat{t_{s}}= \left( \begin{array}{lcc}
     t_{1,s} \\
     \vdots \\
     t_{M-1,s}\\
     t_{M,s}
     \end{array} \right).
     \end{equation}
Using all the previous result, Ec. (\ref{matrizdetrnsferencia1}) is reduced to,
\begin{equation}
\hat{\Delta} \hat{\delta_{s}}= \hat{P}\hat{T} \hat{t_{s}}- \hat{R}\hat{r_{s}}.
\label{matrixequation}
\end{equation}
It can be proven, as is done in appendices B and C, that the whole calculation is thus reduced to solve a set of two linear equations for each channel. The set of equations to be solved is,
\begin{equation}\label{rsstss}
\begin{array}{lcc}
    \left(
 \begin{array}{lcc}
     -1 & g_{s} \\
\\ -e^{-ik_{x}^{s}} & h_{s} 
 \end{array}   
    \right) \left(\begin{array}{lcc}
        r_{ss} \\
        \\ t_{ss}
    \end{array} \right)= \left( \begin{array}{lcc}
        1 \\
        \\ e^{ik_{x}^{s}}
    \end{array} \right)
\end{array}     
\end{equation}    
with
\begin{equation}\label{gshs}
 \left( \begin{array}{lcc}
        g_{s} \\
        \\ h_{s} 
    \end{array} \right)= \prod_{r=0}^{(N/4)+1} \hat{\varsigma}_{r}
     \left( \begin{array}{lcc}
        e^{iNk_{x}^{s}} \\
        \\  e^{i(N+1)k_{x}^{s}}
    \end{array} \right),
\end{equation}
and $\hat{\varsigma}_{r}$ being a $2\times2$ matrix defined in appendix C. Notice that in fact, the trace stability of $\prod_{r=0}^{(N/4)+1} \hat{\varsigma}_{r}$ contains information about the localization of the wave functions and about the  density of states of the nanoribbon \cite{naumisdisorder,naumistracemap,naumisphason,naumistrends}. Finally, using the results obtained in appendices A, B and C, we obtain that the reflectance, $r_{ss}$,  of our system is given by,

\begin{equation}
|r_{ss}|=\left|  \frac{1-(g_s/h_s) e^{ik_{x}^{s}} }{1-(g_s/h_s) e^{-ik_{x}^{s}} } \right|.
\label{reflectance}
\end{equation}
Since the channels do not couple among them, the transmittance is readily found from $|t_{ss}|^{2}=1-|r_{ss}|^{2}$. From the previous result, one can obtain the conductivity ($G$) and the Fano factor ($F$), which in the Landauer-B\"uttiker formalism are given by \cite{shi2012exact},
\begin{equation}
G=\frac{2e^2}{h}\sum_s^{M}|t_{ss}|^2,
\label{conductance}
\end{equation}
and
\begin{equation}
F=\frac{\sum_s |t_{ss}|^2\left(1-|t_{ss}|^2\right)}{\sum_s |t_{ss}|^2}.
\label{fano}
\end{equation}
%

\section{Application to periodic uniaxial strain}
\label{Results}

In this section we apply the previous results to an uniaxial strained AGN. In particular, we consider the case of an uniaxial spatially periodic strain field, namely, we consider that the position of the atoms in the graphene ribbon are shifted from their original position in pristine graphene by a displacement vector given by, 
\begin{equation}
 u(x_i)=\lambda \cos(2\pi\sigma x_i),
\label{strain}
\end{equation}
where $\sigma^{-1}$ is the wavelength and $\lambda$ is the amplitude of the strain field. $x_j$ is the $x$ component of the position of the $j$-th carbon atom in an unstrained AGN. For a strain field as the one in Ec. (\ref{strain}), it can be proved that the hopping parameters, in the small amplitude limit, are given by,
\begin{equation}
 t_i=t_0 \exp{\left\{\lambda [\cos(2\pi\sigma x_{i+1})-\cos(2\pi\sigma x_i]\right\}}.
 \label{hoppingperiodic}
\end{equation}
Since we are dealing with periodic oscillations, the commensurability between the graphene lattice parameter and strain wavelength leads to a superlattice. If both parameters are incommensurate, then a quasicrystal is obtained. However, from here on, we will only consider the case of superlattices.
\begin{figure}
\centering
\includegraphics[scale=0.45]{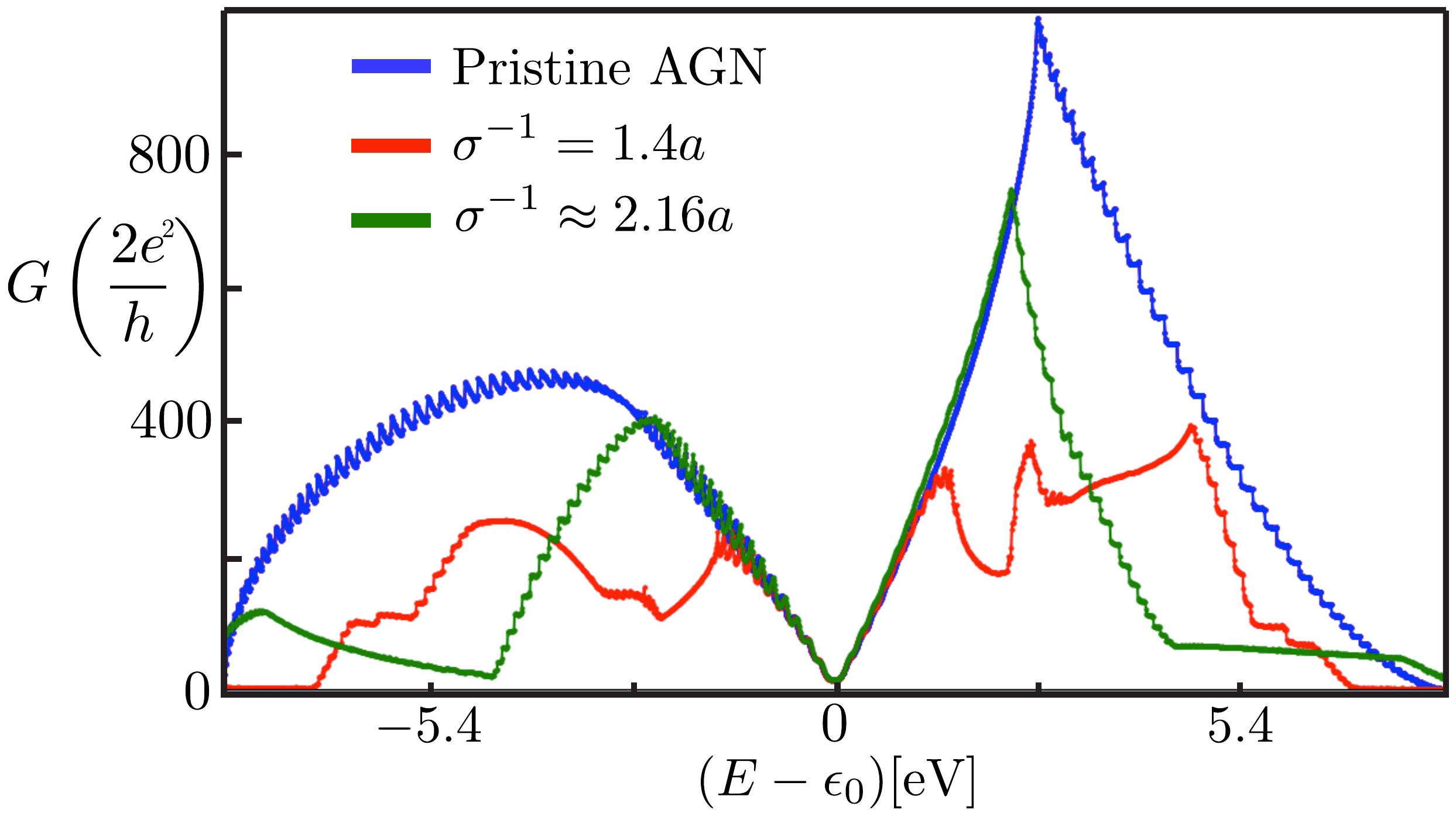}
\caption{conductivity of a strained AGN as a function of $E-\epsilon_0$ for pristine graphene $\sigma=0$ (blue color), 
and a periodic sinusoidal strain with $\sigma=5/7$ (red color), and $\sigma=(3*89)/(4*144)$ (green color). All the curves were obtained for $E=0$, $\lambda=0.04$, $M=1000$, and $N=100$.}
\label{Ocond}
\end{figure}

Now that we have introduced the strain field to be studied, we proceed to obtain the transport properties of the strained AGN for different values of the strain's wavelength and amplitude, here denoted by $\sigma^{-1}$ and $\lambda$, respectively. For the sake of simplicity, we will fix the strain's amplitude and study the transport properties of our system for two values of $\sigma$ (inverse strain's wavelength), which are $\sigma=5/7\approx 0.71$ and $\sigma=(3*89)/(4*144)\approx 0.46$. The latter case was chosen to be three quarters of the ratio of two subsequent Fibonacci numbers, which are rational approximates to the irrational golden mean $\left[\left(\sqrt{5}-1\right)/2\right]$.

\begin{figure}
\centering
\includegraphics[scale=0.45]{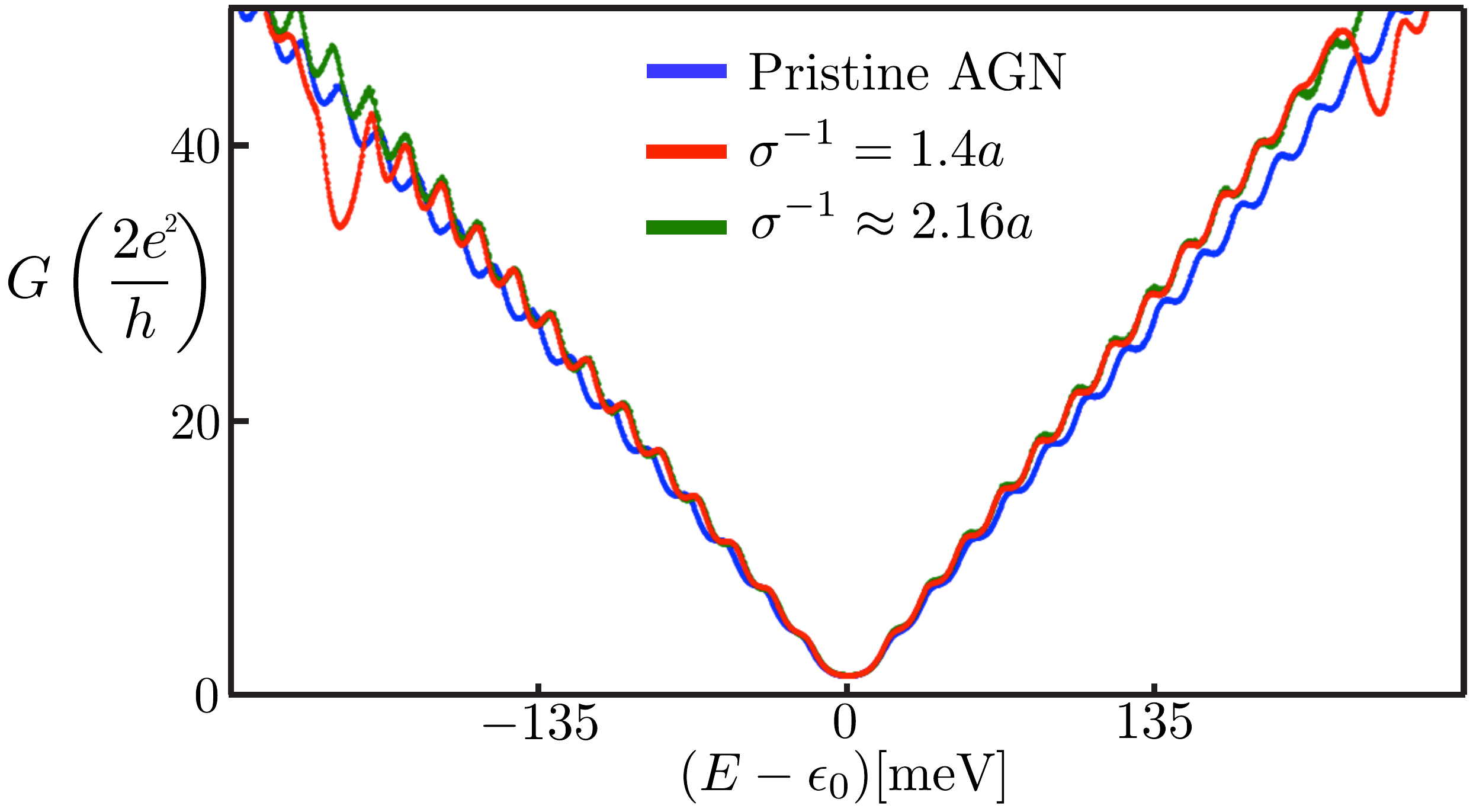}
\caption{Low energy conductivity of a strained AGN, as a function of $E-\epsilon_0$ for different values of the strain's wavelength ($\sigma$). In all cases we have used $E=0$, $\lambda=0.08$, $M=10000$, and $N=1000$.}
\label{LowG}
\end{figure}

\begin{figure}
\centering
\includegraphics[scale=0.46]{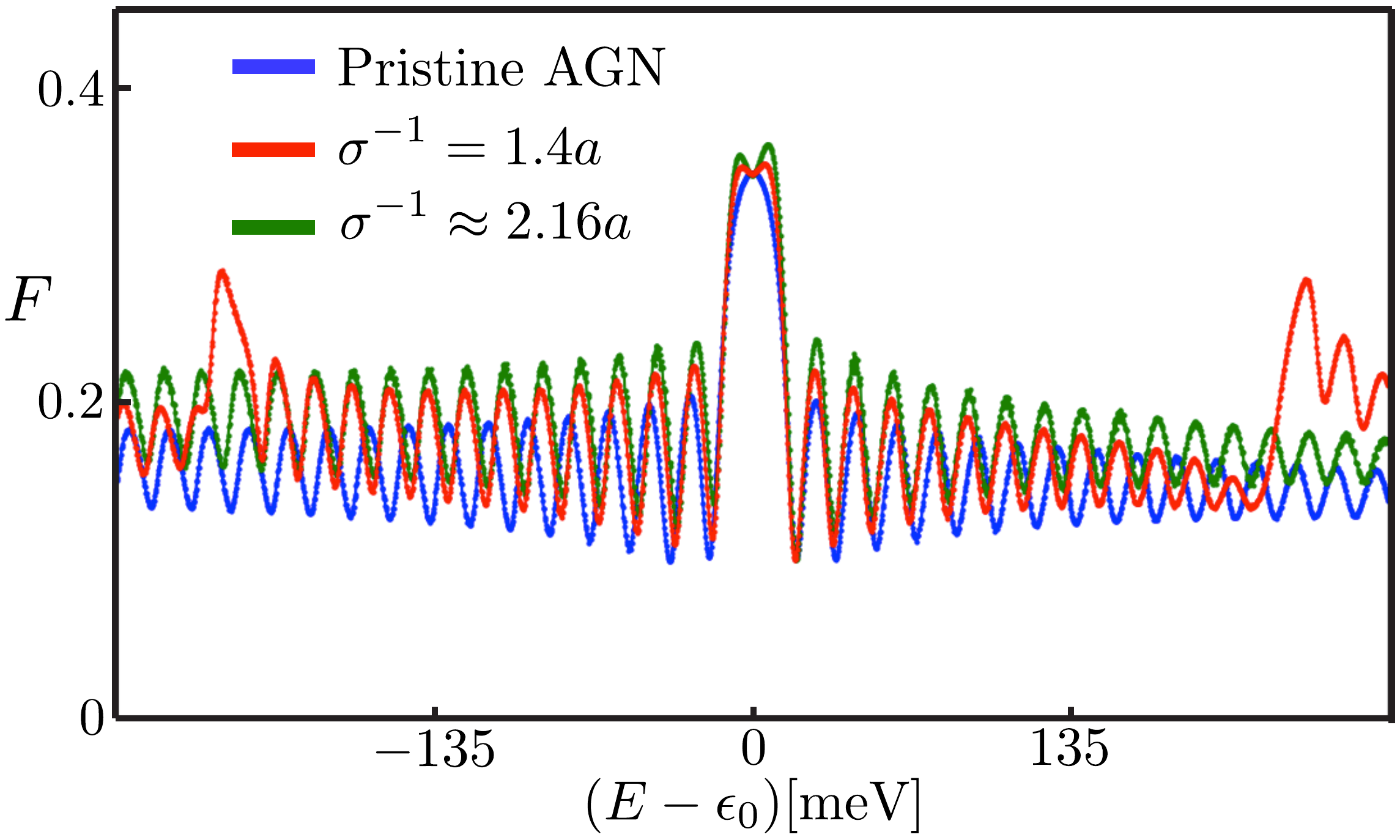}
\caption{Fano factor of a strained AGN, as a function of $E-\epsilon_0$ for different values of the strain's wavelength ($\sigma$). In all cases we have used $E=0$, $\lambda=0.08$, $M=10000$, and $N=1000$.}
\label{FanoFactorFig}
\end{figure}

The case $\sigma=5/7$ corresponds to a system with fourteen different hopping parameters, this is, the system can be seen as a superlattice made from repeating a super cell having fourteen non-equivalent atoms. The case $\sigma=(3*89)/(4*144)$ is an approximation of the quasiperiodic case, using a super cell of $144$ sites. In Fig. \ref{Ocond} we show the conductivity of our system for three different strain's wavelength values, {\it i)} pristine graphene (blue lines), {\it ii)} strained graphene with $\sigma=5/7$ (red lines), and strained graphene with $\sigma=(3*89)/(4*144)$ (green lines). All the plots were made using $N=100$, $M=1000$, and, for the strained cases, $\lambda=0.04$ for Fig. \ref{Ocond}, whereas $N=1000$, $M=10000$, and $\lambda=0.08$ were used for Figs. \ref{LowG} and \ref{FanoFactorFig}. In addition, we have considered the case of half filling for the leads, in other words, we have set the Fermi energy of the electrodes to be $E=0$. The first important observation is that our results reproduce exactly the results found in ref. \cite{shi2012exact}, including the cases of uniform strain (not all presented here).

Many interesting features are observed in Fig. \ref{Ocond}, for example, there is an asymmetry with respect to $E-\epsilon_0=0$ in the conductivity for all the cases. This fact is related to the hole-electron asymmetry induced by the metallic electrodes in the AGN case \cite{zhang2011dependence}. Also, for $\sigma= (3*89)/(4*144) $ we observe that the conductivity is greatly decreased when compared to the pristine case, although it has the same overall behavior. This kind of behavior is quite similar to the one observed for the case in which the AGN is uniformly strained \cite{jing2015mechanical,zenan2015coupling}, as we expect since for $\sigma= (3*89)/(4*144) $ the strain wavelength is $\sigma^{-1}\approx2.16 a$, which is big enough to greatly suppress the effects of the non-uniform strain field. Finally, in Fig. \ref{Ocond}) we present in red solid lines the case $\sigma= 5/7$ (in other words, a wavelength given by $\sigma^{-1}=1.4a$). Note how the conductivity presents peaks as $E-\epsilon_0$ varies. These peaks are a consequence of the difference between the strain wavelength and the periodicity of the unstrained AGN, which leads to dispersion effects \cite{naumis2014mapping,roman2014spectral}. 

Even though there are differences in the conductivities shown in Fig. \ref{Ocond}, they seem to be quite similar around the zero of $E-\epsilon_0$, which indicates that these particular strain fields have no great impact on the transport properties of the AGN at low energies. To confirm this statement, in Figs. \ref{LowG} and \ref{FanoFactorFig} we plot the conductivity and the Fano factor for a much larger AGN in the low energy regime. From the conductivity, in the low energy regime, Fig. \ref{LowG}, we can see that the behavior is basically the same in the three cases displayed therein. For $\sigma=(3*89)/(4*144)$, the effects of the strain field are to shift the oscillations of the conductivity as $E-\epsilon_0$ varies and to increase a little the slope of the conductivity. On the other hand, for $\sigma=5/7$, we also observe a small increase of the conductivity's slope and for energies around $\pm270[\text{meV}]$ the emergence of some fluctuations that are larger than the ones observed in the case of a pristine AGN (see Fig. \ref{LowG}, solid red lines). When it comes to the Fano factor, which is displayed in Fig. \ref{FanoFactorFig}, we observe almost the same features that we find for the conductivity. 

\section{CONCLUDING REMARKS}
\label{conclusions}

In this work we have developed a transfer matrix method that allows to obtain the transport properties of an AGN under any kind of uniaxial strain applied along the armchair direction of the ribbon. We have done this by generalizing the method found in the reference \cite{shi2012exact}. As an application, we have studied the case of a spatially periodic strain field, a sinusoidal strain field to be exact. The conductivity and the Fano factor were obtained for two different strain's wavelength, this was done for a neutral AGN, or, in other words, we considered that the Fermi energy of the leads was zero ($E=0$). We found that for a strain field with wavelength $\sigma^{-1}=2.15a$, for most of the conduction channels, the conductivity is greatly reduced  when compared with the pristine case, although the overall shape of the conductivity is quite similar to its pristine counterpart. For the modulated strain with a smaller supercell  (or, in other words, a smaller strain wavelength given by $\sigma^{-1}=1.4a$), we found that the conductivity oscillates as $E-\epsilon_0$ varies. This can be understood as a dispersion effect induced by the  strain field and the possibility of pseudo-Landau levels \cite{meng2013strain-induced,castro2017raise}. On the other hand, even though the conductivities here studied are very different in the whole spectrum, we found that their behavior near the Dirac point energy is almost the same for all the strain fields that we have discussed. This fact was confirmed by obtaining the conductivity and the Fano factor near the neutrality point. Therein, the effect of the strain field is almost negligible, in other words, we just observed a vary small variation on the slope of the conductivity of a strained AGN compared with a pristine AGN. Although this result seems to be quite unexpected, in fact it can be explained by the strong topological protection near the Dirac cones \cite{katsnelson2012graphene,naumis2017electronic}.  

\section*{ACKNOWLEDGMENTS}

This work was supported by DGAPA-PAPIIT project 102717. P. R.-T. acknowledges financial support from Consejo Nacional de Ciencia y Tecnolog\'ia (CONACYT) (M\'exico).

\appendix{}

\section{}

In this appendix we derive Eqs. (\ref{matricesA}), (\ref{matricesB}), (\ref{matricesC}), and (\ref{matricesD}). We start by studying each case out of the four possible for Eq. (\ref{alphafour}).

\textbf{Case 1.} For this case we have $i=4r-3$, where $r=1,...,N/4$, which leads to the following equation,
\begin{equation} \label{alphaA}
\alpha_{4r-4,j}= \frac{E-\epsilon_{0}}{-t_{4r-4}} \alpha_{i,j} - \frac{t_{4r-3}}{t_{4r-4}} (\alpha_{4r-2,j}+\alpha_{4r-2,j-1}).
\end{equation}
It is convenient to write the previous equation in a matrix form. To that end, we define a $M\times M$ lower-bidiagonal matrix with nonzero elements equal to $1$, namely,
\begin{equation} \label{matrizx}
    \textbf{X}= \left( \begin{array}{lccccc}
         1&0 & 0 & 0&\ldots&0  \\
      \\ 1&1 & 0 & 0& \ldots&0 \\ 
      \\ 0&1 & 1 & 0& \ldots&0 \\
      \\ 0& 0&1& 1& \ldots & 0 \\
       \\ \vdots&\vdots&\vdots&\vdots&\ddots&\vdots \\
       \\ 0&0&0&0&\ldots&1
    \end{array}\right).
\end{equation}
Then, it is straightforward to show that Eq. (\ref{alphaA}) can be written as,
\begin{equation} 
\left(\begin{array}{lcc}
     \vec{\alpha}_{4r-4}  \\
  \\   \vec{\alpha}_{4r-3}
\end{array}\right)= \mathbf{A}_r \left(\begin{array}{lcc}
     \vec{\alpha}_{4r-3}  \\
  \\   \vec{\alpha}_{4r-2}
\end{array}\right)
\label{alphaA}
\end{equation}
where
\begin{equation}
\mathbf{A}_r=\left( \begin{array}{lcc}
  \frac{E-\epsilon_{0}}{-t_{4r-4}} \mathcal{I}_{M\times M }   & \frac{-t_{4r-3}}{t_{4r-4}} \textbf{X} \\
\\    \mathcal{I}_{M \times M} & \textbf{0} 
\end{array}\right).
\label{matrixA}
\end{equation}

\textbf{Case 2.} Here we have $i=4r-2$, with $r=1,...,(N/4)-2$. For this case we obtain,
\begin{equation} \label{alphaB}
\alpha_{4r-1,j}= \frac{E-\epsilon_{0}}{-t_{4r-2}} \alpha_{4r-2,j}- \frac{t_{4r-3}}{t_{4r-2}}(\alpha_{4r-3,j}+\alpha_{4r-3,j+1}).
\end{equation}
Once again, we can rewrite the previous equation in a matrix way, 
\begin{equation} \label{matrixBp}
\left(\begin{array}{lcc}
     \vec{\alpha}_{4r-2}  \\
  \\   \vec{\alpha}_{4r-2}
\end{array}\right)= \left( \begin{array}{lcc}
 \textbf{0}&\mathcal{I}_{M \times M} \\ 
 \\\frac{-t_{4r-3}}{t_{4r-2}} \textbf{X}^{T} & \frac{E-\epsilon_{0}}{-t_{4r-2}} \mathcal{I}_{M\times M }   
\end{array}\right) \left(\begin{array}{lcc}
     \vec{\alpha}_{4r-3}  \\
  \\   \vec{\alpha}_{4r-2}
\end{array}\right)
\end{equation}
where $\textbf{X}^{T}$ is the transpose matrix of $\textbf{X}$. Then one can easily obtain the column vector of the right side of Eq. (\ref{matrixBp}),
\begin{equation}
\left(\begin{array}{lcc}
     \vec{\alpha}_{4r-3}  \\
  \\   \vec{\alpha}_{4r-2}
\end{array}\right)= \mathbf{B}_r\left(\begin{array}{lcc}
     \vec{\alpha}_{4r-3}  \\
  \\   \vec{\alpha}_{4r-1}
\end{array} \right).
\end{equation}
where
\begin{equation}\label{matrixB}
\mathbf{B}_r=\left( \begin{array}{lcc}
\frac{E-\epsilon_{0}}{-t_{4r-3}} \textbf{Y}^{T}  &  \frac{-t_{4r-2}}{t_{4r-3}} \textbf{Y}^{T} \\
\\ \mathcal{I}_{M \times M} & \textbf{0}\\ 
\end{array}\right).
\end{equation}

\textbf{Case 3.} For $i=4r-1$, with $r=1,...,N/4$, we obtain the next equation,
\begin{equation} \label{alphaC}
\alpha_{4r-2,j}= \frac{E-\epsilon_{0}}{-t_{4r-2}} \alpha_{i,j} - \frac{t_{4r-1}}{t_{4r-2}} (\alpha_{4r,j}+\alpha_{4r,j+1}).
\end{equation}
The previous equation can be rewrite as,
\begin{equation} \label{matrixCp}
\left(\begin{array}{lcc}
     \vec{\alpha}_{4r-2}  \\
  \\   \vec{\alpha}_{4r-1}
\end{array}\right)= \mathbf{C}_r \left(\begin{array}{lcc}
     \vec{\alpha}_{4r-1}  \\
  \\   \vec{\alpha}_{4r}
\end{array}\right)
\end{equation}
where
\begin{equation}
\mathbf{C}_r=
\left( \begin{array}{lcc}
  \frac{E-\epsilon_{0}}{-t_{4r-2}} \mathcal{I}_{M\times M }   & \frac{-t_{4r-1}}{t_{4r-2}} \textbf{X}^{T} \\
\\    \mathcal{I}_{M \times M} & \textbf{0} 
\end{array}\right).
\label{matrixC}
\end{equation}

\textbf{Case 4.} For $i=4r$, with $r=1,...,N/4$, we obtain the following result,
\begin{equation} \label{ec10}
\alpha_{4r+1,j}= \frac{E-\epsilon_{0}}{-t_{4r}} \alpha_{4r,j}- \frac{t_{4r-1}}{t_{4r}}(\alpha_{4r-1,j}+\alpha_{4r-1,j-1}).
\end{equation}
As before, if we rewrite the previous equation in a matrix form, we get,
\begin{equation} \label{ec11}
\left(\begin{array}{lcc}
     \vec{\alpha}_{4r}  \\
  \\   \vec{\alpha}_{4r+1}
\end{array}\right)= \left( \begin{array}{lcc}
 \textbf{0}&\mathcal{I}_{M \times M} \\ 
 \\\frac{-t_{4r-1}}{t_{4r}} \textbf{X} & \frac{E-\epsilon_{0}}{-t_{4r}} \mathcal{I}_{M\times M }   
\end{array}\right) \left(\begin{array}{lcc}
     \vec{\alpha}_{4r-1}  \\
  \\   \vec{\alpha}_{4r}
\end{array}\right).
\end{equation}
Note that the Eq. (\ref{ec11}) is quit similar to Eq. (\ref{matrixBp}), therefore, one can proceed as before to find that,
\begin{equation} \label{matrixDp}
\left(\begin{array}{lcc}
     \vec{\alpha}_{i-1}  \\
  \\   \vec{\alpha}_{i}
\end{array}\right)= \mathbf{D}_r
\left(\begin{array}{lcc}
     \vec{\alpha}_{i}  \\
  \\   \vec{\alpha}_{i+1}
\end{array} \right)
\end{equation}
with
\begin{equation}
\mathbf{D}_r=\left( \begin{array}{lcc}
\frac{E-\epsilon_{0}}{-t_{i-1}} \textbf{Y}  &  \frac{-t_{i}}{t_{i-1}} \textbf{Y} \\
\\ \mathcal{I}_{M \times M} & \textbf{0}\\ 
\end{array}\right).
\end{equation}
%

\section{} 

In this appendix we provide the explicit evaluation of the product $\mathbf{A}_r\mathbf{B}_r\mathbf{C}_r\mathbf{D}_r$. Using the definitions of the matrices $\mathbf{A}_r, \ \ \mathbf{B}_r, \ \ \mathbf{C}_r,$ and $\mathbf{D}_r$ given in the previous appendix and after some algebraic manipulations, one gets,
\begin{equation}\label{bloquematrix}
\textbf{A}_{r}\textbf{B}_{r}\textbf{C}_{r}\textbf{D}_{r}=
\left(\begin{array}{lcc}
\vartheta_{(r,1)} \textbf{Y}^{T}\textbf{Y}-\vartheta_{(r,2)} \mathcal{I}_{M \times M}+ \vartheta_{(r,3)} \textbf{XX}^{T}  & \vartheta_{(r,4)} \textbf{Y}^{T}\textbf{Y} - \vartheta_{(r,5)} \mathcal{I}_{M \times M}  \\
\\ \vartheta_{(r,6)} \textbf{Y}^{T}\textbf{Y}+ \vartheta_{(r,7)}\mathcal{I}_{M \times M}     & \vartheta_{(r,8)} \textbf{Y}^{T}\textbf{Y}
\end{array} \right)
\end{equation}
with
\begin{equation}
\vartheta_{(r,1)} =\frac{(E-\epsilon_{0})^{4}- (E-\epsilon_{0})^{2} t_{q_{(r)}+1}^{2}}{t_{q_{(r)}-1} t_{q_{(r)}}t_{q_{(r)}+1}t_{q_{(r)}+2}}
\end{equation}
\begin{equation}
\vartheta_{(r,2)} = \frac{(E-\epsilon_{0})^{2} \left(t_{q_{(r)}+2}^{2}+ t_{q_{(r)}}^{2} \right)}{t_{q_{(r)}-1}t_{q_{(r)}+1}} 
\end{equation}
\begin{equation}
\vartheta_{(r,3)} =\frac{t_{q_{(r)}}t_{q_{(r)}+2}}{t_{q_{(r)}-1} t_{q_{(r)}}t_{q_{(r)}+1}t_{q_{(r)}+2}} 
\end{equation}
\begin{equation}
\vartheta_{(r,4)} = \frac{(E-\epsilon_{0})^{3}(t_{q_{(r)}+3})- (E-\epsilon_{0})t_{q_{(r)}+1}^{2}t_{q_{(r)}+3}}{t_{q_{(r)}-1} t_{q_{(r)}}t_{q_{(r)}+1}t_{q_{(r)}+2}}
\end{equation}
\begin{equation}
\vartheta_{(r,5)} = \frac{(E-\epsilon_{0}) t_{q_{(r)}}t_{q_{(r)}+3} }{t_{q_{(r)}-1} t_{q_{(r)}+1}t_{q_{(r)}+2}} 
\end{equation}
\begin{equation}
\vartheta_{(r,6)}= \frac{-(E-\epsilon_{0})^{3}+ (E-\epsilon_{0})t_{q_{(r)}+1}^{2}}{ t_{q_{(r)}}t_{q_{(r)}+1}t_{q_{(r)}+2}} 
\end{equation}
\begin{equation}
\vartheta_{(r,7)} = \frac{(E-\epsilon_{0})t_{q_{(r)}+2}}{t_{q_{(r)}}t_{q_{(r)}+1}} 
\end{equation}
\begin{equation}
\vartheta_{(r,8)} = \frac{-(E-\epsilon_{0})^{2} t_{q_{(r)}+3}+t_{q_{(r)}+1}^{2} t_{q_{(r)}+3}}{ t_{q_{(r)}}t_{q_{(r)}+1}t_{q_{(r)}+2}}
\end{equation}
where we have used that $q(r)=4r-3$, with $r=1,...,N/4$.

\section{}

In this appendix we derive Eqs. (\ref{rsstss}) and (\ref{gshs}). We start by considering Eq. (\ref{matrixequation}), which connects the wave functions of the lead-nanoribbon junctions,
%
\begin{equation}\label{matrizdetransmisionytransferencia}
\begin{split}
\left( \begin{array}{lcc}
 \hat{\zeta}& 0  \\
    0 & \hat{\zeta} 
\end{array} \right) \left( \begin{array}{lcc}
 \hat{\xi}(x_{0})\\
  \hat{\xi}(x_{1})
\end{array} \right) = \hat{\boldsymbol{P}}
\left( \begin{array}{lcc}
 \hat{\zeta}& 0  \\
    0 & \hat{\zeta} 
\end{array} \right) 
\left( \begin{array}{lcc}
 \hat{\xi}(x_{N})\\
  \hat{\xi}(x_{N+1})
\end{array} \right) \hat{t}_{s} 
-& \left( \begin{array}{lcc}
 \hat{\zeta}& 0  \\
    0 & \hat{\zeta} 
\end{array} \right) \left( \begin{array}{lcc}
 \hat{\xi}(x_{0})^{*}\\
  \hat{\xi}(x_{1})^{*}
\end{array} \right) \hat{r}_{s},
\end{split}
\end{equation}
%
such an equation can be rewritten, by using the inverse of $\hat{\zeta}$, as,
\begin{equation}\label{matrizdetransmisionytransferencia2}
\begin{split}
\left( \begin{array}{lcc}
 \hat{\xi}(x_{0})\\
  \hat{\xi}(x_{1})
\end{array} \right) = 
\left[ \left( \begin{array}{lcc}
 \hat{\zeta}^{-1}& 0  \\
    0 & \hat{\zeta}^{-1} 
\end{array} \right) 
\hat{\boldsymbol{P}}
\left( \begin{array}{lcc}
 \hat{\zeta}& 0  \\
    0 & \hat{\zeta} 
\end{array} \right) \right]
\left( \begin{array}{lcc}
 \hat{\xi}(x_{N})\\
  \hat{\xi}(x_{N+1})
  \end{array} \right) \hat{t}_{s} 
-&  \left( \begin{array}{lcc}
 \hat{\xi}(x_{0})^{*}\\
  \hat{\xi}(x_{1})^{*}
\end{array} \right) \hat{r}_{s}.
\end{split}
\end{equation}
We next observe that the term in square brackets is just an unitary transformation of the matrix $\hat{\boldsymbol{P}}$ to a new one $\hat{\boldsymbol{P}}'$. Using the definition of $\hat{\boldsymbol{P}}$ given by Eq. (\ref{matrizdetrnsferencia1}), it follows that, 
\begin{equation}\label{matrizdetransmisionytransferencia3}
\begin{split}
\hat{\boldsymbol{P}}'=
\left( \begin{array}{lcc}
{\tilde{t}_{2}} \mathcal{I}_{M \times M} & \textbf{0} \\
\\ \textbf{0} &  \mathcal{I}_{M \times M} 
\end{array}\right)
\prod_{r=1}^{N/4} \left( \textbf{A}_{q_{(r)}}\textbf{B}_{q_{(r)}}\textbf{C}_{q_{(r)}}\textbf{D}_{q_{(r)}}\right)^{\prime}
 \left( \begin{array}{lcc}
{\tilde{t}_{1}} \mathcal{I}_{M \times M} & \textbf{0} \\
\\ \textbf{0} &  \mathcal{I}_{M \times M} 
\end{array}\right)\\
\end{split}
\end{equation}
where the product of matrices is written using the unitary transformation,
\begin{equation}\label{EqNewABCD}
(\textbf{A}_{q_{(r)}}\textbf{B}_{q_{(r)}}\textbf{C}_{q_{(r)}}\textbf{D}_{q_{(r)}})'= \left( \begin{array}{lcc}
    \hat{\zeta} &  0 \\
    0 & \hat{\zeta}
\end{array} \right)^{-1} (\textbf{A}_{q_{(r)}}\textbf{B}_{q_{(r)}}\textbf{C}_{q_{(r)}}\textbf{D}_{q_{(r)}})\left( \begin{array}{lcc}
    \hat{\zeta} &  0 \\
    0 & \hat{\zeta}
\end{array} \right)
\end{equation}
By using  Eq. (\ref{bloquematrix}) and after some algebraic manipulations, one gets, 
\begin{equation} \label{transform}
\begin{split}
&(\textbf{A}_{q_{(r)}}\textbf{B}_{q_{(r)}}\textbf{C}_{q_{(r)}}\textbf{D}_{q_{(r)}})'=\\ &\left(\begin{array}{lcc}
\vartheta_{q_{(r)},1} \hat{\zeta}^{-1}\textbf{Y}^{t}\textbf{Y}\hat{\zeta}-\vartheta_{q_{(r)},2} \mathcal{I}_{M \times M}+ \vartheta_{q_{(r)},3} \hat{\zeta}^{-1}\textbf{XX}^{t}\hat{\zeta}  & \vartheta_{q_{(r)},4} \hat{\zeta}^{-1}\textbf{Y}^{t}\textbf{Y}\hat{\zeta}- \vartheta_{q_{(r)},5} \mathcal{I}_{M \times M}  \\
\\ \vartheta_{q_{(r)},6} \hat{\zeta}^{-1}\textbf{Y}^{t}\textbf{Y}\hat{\zeta}+ \vartheta_{q_{(r)},7}\mathcal{I}_{M \times M}     & \vartheta_{q_{(r)},8} \hat{\zeta}^{-1}\textbf{Y}^{t}\textbf{Y}\hat{\zeta}^{-1}
\end{array} \right)
\end{split}
\end{equation}
%
Note that, from the definition of $\mathbf{X}$, Eq. \eqref{matrizx}, we obtain that,
\begin{equation} \label{x y xtranspuesta}
\textbf{X}\textbf{X}^{T}= \left( \begin{array}{lccccc}
         1&1 & 0 & 0&\ldots&0  \\
      \\ 1& 2 & 1 & 0& \ldots&0 \\ 
      \\ 0&1 & 2 & 1& \ldots&0 \\
      \\ 0& 0&1& 2& \ldots & 0 \\
       \\ \vdots&\vdots&\vdots&\vdots&\ddots&\vdots \\
       \\ 0&0&0&0&\ldots&2
    \end{array}\right)
\end{equation}
The previous matrix, $\textbf{X}\textbf{X}^{T}$, is very similar to the Hamiltonian related to a linear chain of M sites with hopping integrals equal to 1 and self-energy equals to 2, except at the first site. On the other hand, matrix $\hat{\zeta}$ can be seen as the matrix with eigen functions of a chain with Hamiltonian $\textbf{X}\textbf{X}^{T}$, therefore, we have that,
%
\begin{equation} \label{prod1}
\hat{\zeta}^{-1} \textbf{XX}^{t} \hat{\zeta}= \left( 
\begin{array}{lccc}
    1+2\cos(k_{y}^{1}) & \ldots & 0 \\
\\    0 & 2+2\cos(k_{y}^{2}) & \ldots& 0\\
\\   \vdots & \vdots & \ddots & \vdots \\
\\ 0& 0 & \ldots & 2+2\cos(k_{y}^{M})\end{array}
\right)
\end{equation}
and, similarly, 
\begin{equation} \label{prod2}
\hat{\zeta}^{-1} \textbf{Y}^{t}\textbf{Y} \hat{\zeta}= \left( 
\begin{array}{lccc}
    1/[1+2\cos(k_{y}^{1})] & \ldots & 0 \\
\\    0 & 1/[2+2\cos(k_{y}^{2})] & \ldots& 0\\
\\   \vdots & \vdots & \ddots & \vdots \\
\\ 0& 0 & \ldots & 1/[2+2\cos(k_{y}^{M})]\end{array}
\right)
\end{equation}
%
Using the previous results to analyze  Eq. (\ref{transform}), one finds that the transverse modes in the AGN are unmixed by the scattering processes, remaining independent and retaining the free-particle dispersion relation $\tilde{\epsilon}_{s}=-2-2\cos(k_{y}^{s})$. Thus, from Eq. (\ref{transform}), for $r=1,2...,N/4$, we define a new $2\times 2$ matrix $\hat{\varsigma}_{r}$ as,
\begin{equation}
 \hat{\varsigma}_{r}=\left(
 \begin{array}{lcc}
     -\vartheta_{{(r,1)}}\tilde{\epsilon}_{s}^{-1}- \vartheta_{{(r,2)}}-\vartheta_{{(r,3)}}\tilde{\epsilon}_{s} & -\vartheta_{{(r,4)}} \tilde{\epsilon}_{s}^{-1}-\vartheta_{{(r,5)}} \\
\\ -\vartheta_{{(r,6)}} \tilde{\epsilon}_{s}^{-1} +\vartheta_{{(r,7)}}&  -\vartheta_{{(r,8)}} \tilde{\epsilon}_{s}^{-1}
 \end{array}   
    \right)
\end{equation}
Now Eq.(\ref{matrizdetransmisionytransferencia2}) can also be written in terms of  $2\times2$ matrices as follows,
\begin{equation}\label{eqtransmisionreduced}
\begin{split}
\left( \begin{array}{lcc}
 \xi(x_{0})\\
  \xi(x_{1})
\end{array} \right) = 
\left( \begin{array}{lcc}
{\tilde{t}_{2}}  & \textbf{0} \\
\\ \textbf{0} &  1\\ 
\end{array}\right)
\prod_{r=1}^{N/4} \hat{\varsigma}_{r}
 \left( \begin{array}{lcc}
{\tilde{t}_{1}}  & \textbf{0} \\
\\ \textbf{0} &  1 
\end{array}\right)
\left( \begin{array}{lcc}
 \xi(x_{N})\\
 \xi(x_{N+1})
  \end{array} \right) t_{ss}
-&  \left( \begin{array}{lcc}
 \xi(x_{0})^{*}\\
  \xi(x_{1})^{*}
\end{array} \right) r_{ss}
\end{split}
\end{equation}
For the sake of simplicity, we define a new $\hat{\varsigma}_{r}$ for $r=0$,  
\begin{equation}\label{eqvar0}
\begin{split}
 \hat{\varsigma}_{0}=\left( \begin{array}{lcc}
{\tilde{t}_{2}}  & \textbf{0} \\
\\ \textbf{0} &  1 
\end{array}\right)\\
\end{split}
\end{equation}
and for $r=(N/4)+1$,
\begin{equation}\label{eqvarN1}
\begin{split}
 \hat{\varsigma}_{(N/4)+1}=\left( \begin{array}{lcc}
{\tilde{t}_{1}}  & \textbf{0} \\
\\ \textbf{0} &  1 
\end{array}\right)\\
\end{split}
\end{equation}
By using the definitions of $ \hat{\varsigma}_{0}$ and  $\hat{\varsigma}_{(N/4)+1}$, a simplified version of Eq. (\ref{eqtransmisionreduced}) can be obtained, 
\begin{equation}\label{eqtransmisionreduced2}
\begin{split}
\left( \begin{array}{lcc}
 \xi(x_{0})\\
  \xi(x_{1})
\end{array} \right) = 
\prod_{r=0}^{(N/4)+1} \hat{\varsigma}_{r}
\left( \begin{array}{lcc}
 \xi(x_{N})\\
 \xi(x_{N+1})
  \end{array} \right) t_{ss}
-&  \left( \begin{array}{lcc}
 \xi(x_{0})^{*}\\
  \xi(x_{1})^{*}
\end{array} \right) r_{ss}
\end{split}
\end{equation}
This is a system of two simultaneous linear equations for the unknown variables $r_{ss}$ and $t_{ss}$. Using the definition for  $ \xi_{N}$ and  
$\xi_{N+1}$ in the previous expressions, and defining a new vector  that contains the product of matrices, 
\begin{equation}\label{Eq:EffSystem}
\begin{array}{lcc}  
\\ \left(\begin{array}{lcc}
        g_{s} \\
        \\ h_{s} 
    \end{array} \right)= \prod_{r=0}^{(N/4)+1} \hat{\varsigma}_{r}
     \left( \begin{array}{lcc}
        e^{iNk_{x}^{s}} \\
        \\  e^{i(N+1)k_{x}^{s}}
    \end{array} \right)
\end{array}
\end{equation}
from where it follows that,
\begin{equation}\label{Eq:EffSystem}
\begin{array}{lcc}
    \left(
 \begin{array}{lcc}
     -1 & g_{s} \\
\\ -e^{-ik_{x}^{s}} & h_{s} 
 \end{array}   
    \right) \left(\begin{array}{lcc}
        r_{ss} \\
        \\ t_{ss}
    \end{array} \right)= \left( \begin{array}{lcc}
        1 \\
        \\ e^{ik_{x}^{s}}
    \end{array} \right) \\
\end{array}
\end{equation}
and
\begin{equation}
\begin{array}{lcc}
\\ \left( \begin{array}{lcc}
        r_{ss} \\
        \\ t_{ss} 
    \end{array} \right)=    \left(
 \begin{array}{lcc}
     -1 & g_{s} \\
\\ -e^{-ik_{x}^{s}} & h_{s} 
 \end{array}   
    \right)^{-1}\left( \begin{array}{lcc}
        1 \\
        \\ e^{ik_{x}^{s}}
    \end{array} \right)
\end{array}
\end{equation}
%

\bibliographystyle{unsrt}
\bibliography{biblio}
\end{document}